\documentclass[12pt]{article}
\usepackage{epsfig}
\begin{document}
\baselineskip=16pt
\begin{titlepage}
\setcounter{page}{0}
\begin{center}

\vspace{0.5cm}
 {\Large \bf Cosmological Scaling Solutions of Multiple Tachyon Fields
 with Inverse Square Potentials}\\
\vspace{10mm}
Zong-Kuan Guo\footnote{e-mail address: guozk@itp.ac.cn}$^{b}$
and
Yuan-Zhong Zhang$^{a,b}$ \\
\vspace{6mm} {\footnotesize{\it
  $^a$CCAST (World Lab.), P.O. Box 8730, Beijing 100080\\
  $^b$Institute of Theoretical Physics, Chinese Academy of Sciences,
      P.O. Box 2735, Beijing 100080, China\\}}

\vspace*{5mm} \normalsize
\smallskip
\medskip
\smallskip
\end{center}
\vskip0.6in 
\centerline{\large\bf Abstract}
{We investigate cosmological dynamics of multiple tachyon fields with
inverse square potentials. A phase-space analysis of the spatially flat
FRW models shows that there exists power-law cosmological scaling
solutions. We study the stability of the solutions and find that the
potential-kinetic-scaling solution is a global attractor. However, in the
presence of a barotropic fluid the solution is an attractor only in one
region of the parameter space and the tracking solution is an attractor
in the other region. We briefly discuss the physical consequences of
these results.}

\vspace*{2mm}
\end{titlepage}

\section{Introduction}

Cosmological inflation has become an integral part of the standard model
of the universe. Apart from being capable of removing the shortcomings
of the standard cosmology, the paradigm seems to have gained a fairly
good amount of support from the recent observations on microwave
background radiation. On the other hand, there have been difficulties in
obtaining accelerated expansion from fundamental theories such as
string/M-theory. Recently, Sen~\cite{SEN} has constructed a classical
time-dependent solution which describes the decay process of an unstable
D-brane in the open string theory. During the decay process the tachyon
field on the brane rolls down toward the minimum of the potential. The
tachyon field might be responsible for cosmological inflation at the early
epochs due to tachyon condensation near the top of the effective
potential~\cite{GWG}, and could contribute to some new form of
cosmological dark matter at late times. Several authors have investigated
the process of rolling of the tachyon in the cosmological
background~\cite{SCQ} and in the braneworld scenario~\cite{BBS}.

The stability of tachyon inflation against changes in initial conditions
has been studied for exponential potentials~\cite{GPCZ} and for
inverse power-law potentials~\cite{AF}. Exact solutions for a purely tachyon
field with an inverse square potential are known~\cite{TP}, but no
solutions exist for multiple tachyon fields, so a dynamical analysis
may be relevant. It is interesting that the inverse square potentials
play the same role for tachyon fields as the exponential potentials do
for standard scalar fields~\cite{HW}. The potentials allow constructing
an autonomous system which gives power-law solutions.
Recently, it was pointed out that the tachyon model of inflation is hard to
be consistent with observations~\cite{KL}, unless there is a large number
of coincident branes~\cite{PCZZ}. In this letter, we will consider
cosmological dynamics of multiple tachyon fields with inverse
square potentials. We have assumed that there is no direct coupling
between the inverse square potentials. The only interaction is gravitational.
A phase-space analysis of the spatially flat FRW models shows that there
exist a cosmological scaling solution which is a unique attractor.
We find that accelerated expansion of the universe can be driven by multiple
tachyon fields at lower-Planck energy densities. In this model we then
introduce a barotropic perfect fluid to the system. We discuss the physical
consequences of these results.

\section{Multiple Inverse Square Potentials}

We start with more general model with $m$ tachyon fields $\phi _i$,
in which each has an inverse square potential
\begin{equation}
V_i(\phi _i)=V_{0i}\phi_i^{-2}.
\end{equation}
Note that there is no direct coupling of the tachyon fields, which influence
each other only via their effect on the expansion. The evolution equation
of each tachyon field for a spatially flat FRW model with Hubble parameter
$H$ is
\begin{equation}
\label{EE}
\frac{\ddot{\phi_i}}{1-\dot{\phi_i}^2}+3H\dot{\phi_i}+\frac{1}{V_i(\phi_i)}
\frac{dV_i(\phi_i)}{d\phi_i}=0,
\end{equation}
subject to the Friedmann constraint
\begin{equation}
\label{FE}
H^2=\frac{\kappa^2}{3} \sum_{i=1}^m \frac{V_i(\phi_i)}
{\sqrt{1-\dot{\phi_i}^2}},
\end{equation}
where $\kappa ^2 \equiv 8 \pi G_N$ is the gravitational coupling. Defining
$2m$ dimensionless variables
\begin{eqnarray}
x_i &\equiv& \dot{\phi_i}, \\
y_i &\equiv& \frac{\kappa^2 V_i(\phi_i)}{3H^2},
\end{eqnarray}
the evolution equations (\ref{EE}) can be written as an autonomous system:
\begin{eqnarray}
\label{EE1}
x'_i &=& -3\left(x_i-\sqrt{\beta_i y_i}\right)\left(1-x_i^2\right), \\
y'_i &=& 3y_i\left(\sum_{i=1}^m\frac{y_i \, x_i^2}{\sqrt{1-x_i^2}}
-\sqrt{\beta_i y_i}\,x_i\right),
\label{EE2}
\end{eqnarray}
where a prime denotes a derivative with respect to the logarithm of the scalar
factor, $N \equiv \ln a$, and $\beta_i \equiv 4/ (3\kappa^2V_{0i})$.
The constraint equation (\ref{FE}) becomes 
\begin{equation}
\label{CE}
1=\sum_{i=1}^m \frac{y_i}{\sqrt{1-x_i^2}}.
\end{equation}
Critical points correspond to fixed points where $x'_i=0$ and $y'_i=0$, and
there are self-similar solutions with
\begin{equation}
\frac{\dot{H}}{H^2}=-\frac{3}{2}\sum_{i=1}^{m}\frac{y_i \, x_i^2}{\sqrt{1-x_i^2}}.
\end{equation}
This corresponds to an expanding universe with a scale factor $a(t)$ given
by $a\propto t^p$ or a contracting universe with a scalar factor given by
$a \propto (-t)^p$, where
\begin{equation}
\label{SFP}
p=\frac{1}{\frac{3}{2}\sum_{i=1}^{m}\frac{y_i \, x_i^2}{\sqrt{1-x_i^2}}}.
\end{equation}

Setting $x'_i=0$ and $y'_i=0$, we can get the fixed points $K$ corresponding
to kinetic-dominated solutions and the fixed points $S$ listed in Table 1 with
\begin{equation}
\label{GS}
x_s^2=\frac{1}{2}\beta \left(\sqrt{\beta^2+4}-\beta\right),
\end{equation}
where we using the following definition:
\begin{equation}
\frac{1}{\beta} \equiv \sum_{i=1}^{m}\frac{1}{\beta_i}.
\end{equation}
The fixed points $K$ behave like non-relativistic matter with $p=2/3$.
The fixed point $S$ is the kinetic-potential-scaling solution with
\begin{equation}
\label{PP}
p=\frac{4}{3\beta (\sqrt{\beta^2+4}-\beta)}.
\end{equation}
We find that accelerated expansion of the universe occurs if $\beta < 2/\sqrt{3}$.
For a single tachyon field with an inverse square potential, $V_0$ must
be larger than $2/(\sqrt{3}\kappa^2)$ to guarantee accelerated expansion.
However, accelerated expansion of the universe can be driven by sufficiently
multiple tachyon fields with $V_0$ at lower-Planck energy densities.

\begin{table}
\begin{tabular}{c c c c c c} \hline
Label & $x_i$ & $y_i$ & $p$ & Existence & Stability \\ \hline
$K$ & $\pm 1$ & 0 & $\frac{2}{3}$ & all $\beta_i$
    & unstable \\
$S$ & $x_s$ & $\frac{x_s^2}{\beta_i}$
 & $\frac{4}{3\beta(\sqrt{\beta^2+4}-\beta)}$ & all $\beta_i$
 & stable \\ \hline
\end{tabular}
\caption{The properties of the critical points in a spatially flat FRW
universe containing two tachyon fields with inverse square potentials.}
\end{table}

In order to analysis the stability of the critical points, we only consider the
cosmologies containing two tachyon fields. Using the Friedmann constraint
equation (\ref{CE}), we reduce Eqs.(\ref{EE1}) and (\ref{EE2}) to three
independent equations
\begin{eqnarray}
\label{IE1}
x'_1 &=& -3\left(x_1-\sqrt{\beta_1 y_1}\right)\left(1-x_1^2\right), \\
y'_1 &=& 3y_1\left[\frac{(x_1^2-x_2^2)y_1}{\sqrt{1-x_1^2}}+x_2^2
 -\sqrt{\beta_1 y_1}\, x_1 \right], \\
x'_2 &=& -3\left[x_2-\sqrt{\beta_2}\left(1-x_2^2\right)^{\frac{1}{4}}
 \left(1-\frac{y_1}{\sqrt{1-x_1^2}}\right)^{\frac{1}{2}}\right]\left(1-x_2^2\right).
\label{IE3}
\end{eqnarray}
The linearization of system (\ref{IE1})-(\ref{IE3}) about the fixed points $K$
yields three positive characteristic exponents
\begin{displaymath}
\lambda_1=\lambda_2=6, \;
\lambda_3=3,
\end{displaymath}
which indicate that the kinetic-dominated solutions are always unstable.
Substituting linear perturbations about the critical point $S$ into
Eqs.(\ref{IE1})-(\ref{IE3}), to first-order in the perturbations,
yields the three negative eigenvalues
\begin{eqnarray}
\lambda_1 &=& -\frac{3x_s^2}{2\beta^2}(2x_s^2+\beta^2), \nonumber \\
\lambda_2 &=& -\frac{3x_s^2}{4\beta^2}\left[(2x_s^2+\beta^2)
 +\sqrt{(2x_s^2+\beta^2)^2
 -16\beta^2x_s^2(1+\beta_2/\beta_1)(1-\beta/\beta_1)} \right], \nonumber \\
\lambda_3 &=& -\frac{3x_s^2}{4\beta^2}\left[(2x_s^2+\beta^2)
 -\sqrt{(2x_s^2+\beta^2)^2
 -16\beta^2x_s^2(1+\beta_2/\beta_1)(1-\beta/\beta_1)} \right]. \nonumber
\end{eqnarray}
The critical point is consequently stable so that the corresponding
cosmological scaling solution is always a global attractor for any
$\beta$. For a single tachyon field model $\beta=\beta_1$, the three
eigenvalues reduce to one eigenvalue.

\section{Plus a Barotropic Perfect Fluid}

We now consider multiple tachyon fields with inverse square potentials
evolving in a spatially flat FRW universe containing a fluid with barotropic
perfect equation of state $P_\gamma=(\gamma-1)\rho_\gamma$, where
$\gamma$ is a constant, $0<\gamma\le2$, such as radiation ($\gamma=4/3$)
or dust ($\gamma=1$). The evolution equation for the
barotropic perfect fluid is
\begin{equation}
\label{BE}
\dot{\rho_ \gamma}=-3H(\rho_\gamma+P_\gamma),
\end{equation}
subject to the Fridemann constraint
\begin{equation}
H^2=\frac{\kappa ^2}{3}\left(\sum_{i=1}^{m}\frac{V_i(\phi_i)}
{\sqrt{1-\dot{\phi_i}^2}}+\rho_\gamma \right).
\end{equation}
We define another dimensionless variable
$z \equiv \frac{\kappa^2 \rho_\gamma}{3H^2}$. The evolution equations
(\ref{EE}) and (\ref{BE}) can then be written as an autonomous system:
\begin{eqnarray}
\label{SY1}
x'_i &=& -3\left(x_i-\sqrt{\beta_i y_i}\right)\left(1-x_i^2\right), \\
y'_i &=& 3y_i\left(\sum_{i=1}^m\frac{y_i \, x_i^2}{\sqrt{1-x_i^2}}
 -\sqrt{\beta_i y_i}\,x_i +\gamma z \right), \\
z'   &=& 3z\left(\sum_{i=1}^m\frac{y_i \, x_i^2}{\sqrt{1-x_i^2}}
 +\gamma z-\gamma \right),
\label{SY2}
\end{eqnarray}
and the Fridemann constraint equation becomes
\begin{equation}
1=\sum_{i=1}^m \frac{y_i}{\sqrt{1-x_i^2}} + z.
\end{equation}
Critical points correspond to fixed points where $x'_i=0$, $y'=0$ and
$z'=0$, and there are self-similar solutions with
\begin{equation} 
\frac{\dot{H}}{H^2}=-\frac{3}{2}\left(\sum_{i=1}^{m}\frac{y_i \, x_i^2}
 {\sqrt{1-x_i^2}} + \gamma z \right).
\end{equation}
This corresponds to an expanding universe with a scale factor $a(t)$ given
by $a\propto t^p$, where 
\begin{equation}
p=\frac{2}{3\sum_{i=1}^{m}\frac{y_i \, x_i^2}{\sqrt{1-x_i^2}}+3\gamma z}\,.
\end{equation}
The system (\ref{SY1})-(\ref{SY2}) has a fixed point $K$ corresponding to 
kinetic-dominated solution, a fixed point $S$ which is a
kinetic-potential-scaling solution, a fixed point $F$ which is a
fluid-dominated solution, and a fixed point $T$ which is a tracking
solution listed in Table 2. In order to analysis the stability of the critical
points, we still consider the cosmologies containing two tachyon fields
plus a barotropic perfect fluid.

\begin{table}
\begin{tabular}{c c c c c c c} \hline
Label & $x_i$ & $y_i$ & $z$ & $p$ & Existence & Stability \\ \hline
$K$ & $\pm 1$ & 0 & 0 & $\frac{2}{3}$ & all $\beta_i, \gamma$
    & unstable \\
$S$ & $x_s$ & $\frac{x_s^2}{\beta_i}$ & 0
 & $\frac{4}{3\beta(\sqrt{\beta^2+4}-\beta)}$ & all $\beta_i, \gamma$
 & stable ($\gamma \ge \frac{\beta\sqrt{\beta^2+4}-\beta^2}{2}$); \\
 &&&&&& unstable ($\gamma < \frac{\beta\sqrt{\beta^2+4}-\beta^2}{2}$) \\
$F$ & 0 & 0 & 1 & $\frac{2}{3\gamma}$ & all $\beta_i, \gamma$
 & stable ($\gamma = 0$); \\
 &&&&&& unstable ($\gamma \ne 0$) \\
$T$ & $\sqrt{\gamma}$ & $\frac{\gamma}{\beta_i}$
 & $\frac{\beta \sqrt{1-\gamma}-\gamma}{\beta \sqrt{1-\gamma}}$
 & $\frac{2}{3\gamma}$ & $\gamma<\frac{\beta\sqrt{\beta^2+4}-\beta^2}{2}$
 & stable \\ \hline
\end{tabular}
\caption{The properties of the critical points in a spatially flat FRW
universe containing two tachyon fields with inverse square
potentials plus a barotropic perfect fluid.}
\end{table}

$K$: $x_i=\pm 1$, $y=0$, $z=0$. These kinetic-dominated solutions
always exist for all $\beta_i$ and $\gamma$, which behave like
non-relativistic matter with $a\propto t^{2/3}$ irrespective of the nature
of the potentials. The linearization of system (\ref{SY1})-(\ref{SY2})
about these fixed points yields four eigenvalues
\begin{displaymath}
\lambda_1=\lambda_2=6, \;
\lambda_3=\lambda_4=3\gamma,
\end{displaymath}
which indicate that the solutions are always unstable.

$S$: $x_i=x_s$, $y=x_s^2/\beta_i$, $z=0$. The potential-kinetic-scaling
solution exists for all $\beta_i$ and $\gamma$. The power-law exponent
(\ref{PP}) depends on the parameter $\beta$ of the potentials. The
linearization of system (\ref{SY1})-(\ref{SY2}) about the fixed point yields
four eigenvalues
\begin{eqnarray}
\lambda_1 &=& -3\left(\frac{x_s^4}{\beta^2}+\frac{x_s^2}{2}\right), \nonumber \\
\lambda_2 &=& -3(\gamma-x_s^2), \nonumber \\
\lambda_3 &=& -\frac{3}{2}\left(\frac{x_s^4}{\beta^2}+\frac{x_s^2}{2}\right)
 +\frac{3}{2}\sqrt{\left(\frac{x_s^4}{\beta^2}+\frac{x_s^2}{2}\right)^2
 -\frac{4x_s^6}{\beta^2}}, \nonumber \\
\lambda_4 &=& -\frac{3}{2}\left(\frac{x_s^4}{\beta^2}+\frac{x_s^2}{2}\right)
 -\frac{3}{2}\sqrt{\left(\frac{x_s^4}{\beta^2}+\frac{x_s^2}{2}\right)^2
 -\frac{4x_s^6}{\beta^2}}, \nonumber
\end{eqnarray}
which indicate that the solutions are unstable for
$\gamma < \beta(\sqrt{\beta^2+4}-\beta)/2$ and stable for
$\gamma \ge \beta(\sqrt{\beta^2+4}-\beta)/2$ from Eq.(\ref{GS}).

$F$: $x_i=0$, $y_i=0$, $z=1$. The fluid-dominated solution exists for
all $\beta_i$ and $\gamma$, corresponding to a power-law solution with
$p=2/3\gamma$. The linearization of system (\ref{SY1})-(\ref{SY2}) about
the fixed point yields four eigenvalues
\begin{displaymath}
\lambda_1=\lambda_2=-3, \;
\lambda_3=\lambda_4=3\gamma,
\end{displaymath}
which indicate that the solution is unstable for $\gamma \ne 0$
and stable only for $\gamma = 0$.

$T$: $x_i=\sqrt{\gamma}$, $y_i=\gamma/\beta_i$,
$z=1-\frac{\gamma}{\beta \sqrt{1-\gamma}}$. The tracking solution exists
for $\gamma<\beta(\sqrt{\beta^2+4}-\beta)/2$. The solution displays a
tracking behavior according to the definition in~\cite{LS}. The power-law
exponent, $p=2/3\gamma$, is identical to that of the fluid-dominated
solution, depends only on the barotropic index $\gamma$ and is
independent of the parameters $\beta_i$ of the potentials. The
linearization of system (\ref{SY1})-(\ref{SY2}) about the fixed point
yields four eigenvalues
\begin{eqnarray}
\lambda_1 &=& -\frac{3}{4}(2-\gamma)
 +\frac{3}{4}\sqrt{4-20\gamma+17\gamma^2}, \nonumber \\
\lambda_2 &=& -\frac{3}{4}(2-\gamma)
 -\frac{3}{4}\sqrt{4-20\gamma+17\gamma^2}, \nonumber \\
\lambda_3 &=& -\frac{3}{4}(2-\gamma)
 +\frac{3}{4}\sqrt{(2-\gamma)^2+16\gamma\sqrt{1-\gamma}
 \left(\gamma/\beta-\sqrt{1-\gamma}\right)}, \nonumber \\
\lambda_4 &=& -\frac{3}{4}(2-\gamma)
 -\frac{3}{4}\sqrt{(2-\gamma)^2+16\gamma\sqrt{1-\gamma}
 \left(\gamma/\beta-\sqrt{1-\gamma}\right)}, \nonumber
\end{eqnarray}
which indicate that the solution is always stable when this point exists for
$\gamma<\beta(\sqrt{\beta^2+4}-\beta)/2$.

The different regions in the ($\beta$, $\gamma$) parameter space lead
to different qualitative evolution in Figure 1.
For $\gamma>\beta(\sqrt{\beta^2+4}-\beta)/2$, $K$, $S$ and $F$ exist.
Point $S$ is the stable late-time attractor. Hence
generic solutions begin at the two kinetic-dominated solutions or at the
fluid-dominated solution and approach the kinetic-potential-scaling solution at
late times. For $\gamma<\beta(\sqrt{\beta^2+4}-\beta)/2$, all critical points exist.
Point $T$ is the stable late-time attractor. Hence generic solutions start at the
two kinetic-dominated solution, at the kinetic-potential-scaling solution or at the
fluid-dominated solution and approach the stable fluid-kinetic-potential-scaling
solution.

\begin{figure}
\includegraphics[scale=1.6]{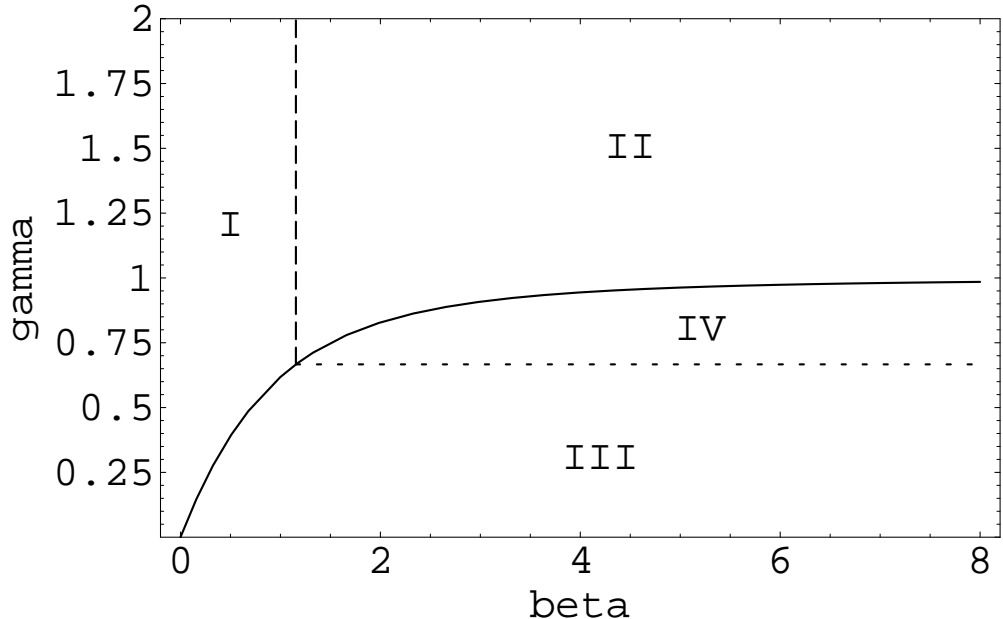}
\caption{Stability regions of the ($\beta$, $\gamma$) parameter space.
In the regions I and II, the tachyon kinetic-potential scaling solution is
the stable late-time attractor. In the regions III and IV, the fluid-dominated
solution is the stable late-time attractor. The universe accelerates in the
regions I and III, while the universe decelerates in the regions II and IV.}
\end{figure}

\section{Conclusions and Discussions}

We have presented a phase-space analysis of the evolution for a spatially
flat FRW universe containing $m$ tachyon fields with inverse square
potentials. We find that there exist cosmological scaling solutions that
corresponds to an expanding universe with $a\propto t^p$. For a single
tachyon field with an inverse square potential, the universe could accelerate
only at nearly Planckian energy densities. However, accelerated expansion
of the universe can be driven by sufficiently multiple tachyon fields
even at lower-Planck energy densities. The reason for this behavior is
that while each field experiences the `downhill' force from its own
potential, it feels the friction from all the tachyon fields via their
contribution to the expansion~\cite{LM}. In order to analysis the stability
of the critical points, we only consider the cosmologies containing two
tachyon fields. We find that the critical point is always stable so that
the scaling solution is a global attractor irrespective of the form of
the potentials. This implies that the velocity of each tachyon field tends
to be equal and constant via their effect on the expansion.

Then we have extended the phase-space analysis of the evolution to a
realistic universe model with a barotropic perfect fluid plus $m$ tachyon
fields with inverse square potentials. We have shown that the energy
density of the tachyon dominates at late times for
$\gamma>\beta(\sqrt{\beta^2+4}-\beta)/2$, In constraint, for
$\gamma<\beta(\sqrt{\beta^2+4}-\beta)/2$, the barotropic fluid does not
dominate completely and the contribution of tachyon energy density to
the total one is not negligible.

We emphasize that we have assumed that there is no direct coupling
between these inverse square potentials. In a system with $m$
non-coincident but parallel non-BPS D3-branes~\cite{MPP},
there are two kinds of open strings. One of them starts
from and ends on the same brane; The other starts from a given brane,
then ends on a different brane. If the distance between two branes are
much larger than the string scale, one can ignore the second kind of
open strings, leaving a tachyon on the world volume for every brane.
Thus, we have $m$ tachyons without interaction and the action is simply
the sum of $m$ single-tachyon actions. It is worth studying further the
the cosmological dynamics of multiple tachyon fields with interactions.

\section*{Acknowledgements}

It is a pleasure to acknowledge helpful discussions with Yun-Song Piao
and Rong-Gen Cai. This project was in part supported by
NBRPC2003CB716300.

\end{document}